\documentclass[12pt]{article}

\usepackage{amsmath}
\usepackage{newtxtext, newtxmath}

\usepackage{graphicx}

\usepackage{geometry}
\usepackage{cite}

\usepackage{indentfirst}

\usepackage{hyperref}

\def\vector#1{\mbox{\boldmath $#1$}}

\title{Axial momentum gains of ions and electrons \\ in magnetic nozzle acceleration}
\author{Kazuma Emoto,\textsuperscript{1,a)} Kazunori Takahashi,\textsuperscript{2} and Yoshinori Takao\textsuperscript{3,b)}}
\date{}

\begin{document}

\maketitle

\noindent
\textsuperscript{1}\textit{Department of Mechanical Engineering, Materials Science, and Ocean Engineering, Yokohama National University, Yokohama 240-8501, Japan}

\noindent
\textsuperscript{2}\textit{Department of Electrical Engineering, Tohoku University, Sendai 980-8579, Japan}

\noindent
\textsuperscript{3}\textit{Division of Systems Research, Yokohama National University, Yokohama 240-8501, Japan}

\noindent
\textsuperscript{a)}E-mail: kazuma-emoto-vh@ynu.jp

\noindent
\textsuperscript{b)}E-mail: takao@ynu.ac.jp

\section*{Abstract}

The fully kinetic simulations of magnetic nozzle acceleration are conducted to investigate the axial momentum gains of ions and electrons with the electrostatic and Lorentz forces. Axial momentum gains per ion and electron are directly calculated from the kinetics of charged particles, indicating that electrons in the magnetic nozzle obtain the net axial momentum by the Lorentz force even though they are decelerated by the electrostatic force. Whereas ions are also accelerated by the electrostatic force, the axial momentum gain of electrons increases significantly with increasing the magnetic field strength and becomes dominant in the magnetic nozzle. In addition, it is clearly shown that the axial momentum gain of electrons is due to the electron momentum conversion from the radial to axial direction, resulting in the significant increase in the thrust and the exhaust velocity. 

\section{Introduction}
\label{sec:introduction}

Electric propulsion systems are recognized as important devices to carry out space missions because of the advantage of the high specific impulse \cite{Goebel2008_fundamentals, Charles2009_jpd, Mazouffre2016_psst, Levchenko2020_pop}. In electric propulsion systems, ion thrusters and Hall thrusters are successfully operated in space \cite{Racca1998_eb, Kuninaka2007_jpp, Brophy2009_jpp}. However, those lifetime is limited by the wear of cathodes and neutralizers since they are damaged by ion sputtering \cite{Kim1998_jpp, Goebel2011_jpp}. To overcome the problem of the electrode wear and extend the lifetime of electric propulsion systems, electrodeless plasma thrusters have been developed vigorously, which do not require cathodes and the neutralizer, e.g., variable specific impulse plasma rocket \cite{Arefiev2004_pop}, helicon double layer thrusters \cite{Charles2007_psst}, and magnetic nozzle radiofrequency (rf) plasma thrusters \cite{Takahashi2019_rmpp}. In particular, the performance of the magnetic nozzle rf plasma thruster has been increased significantly in recent years \cite{Takahashi2021_sr}.

The physics of the magnetic nozzle acceleration has been investigated by experiments, analytical models, and numerical simulations \cite{Little2013_pop, Lafleur2014_pop, Williams2015_ieee, Takahashi2019_rmpp, Kaganovich2020_pop}. Previous studies showed that a diamagnetic effect induced the azimuthal drift current in the magnetic nozzle, which produced the Lorentz force and gave the thrust to the thruster system \cite{Ahedo2010_pop, Takahashi2011_prl, Fruchtman2012_pop, Takahashi2013_prl, Takahashi2016_psst}. More recently, the effects of drift currents in the magnetic nozzle were analyzed using the fully kinetic simulation, indicating that the main drift that produced the Lorentz force was the diamagnetic effect with the strong magnetic field strength \cite{Emoto2021_pop}.

The forces exerted on ions $\vector{f}_\mathrm{i}$ and electrons $\vector{f}_\mathrm{e}$ in the electromagnetic propulsion system are generally written as

\begin{equation}
    \vector{f}_\mathrm{i} = q n_\mathrm{i} \vector{E},
\end{equation}

\begin{equation}
    \vector{f}_\mathrm{e} = - e n_\mathrm{e} \vector{E} + \vector{j}_\mathrm{e} \times \vector{B},
\end{equation}

\noindent
where $q$ is the charge of ions, $e$ is the elementary charge, $n_\mathrm{i}$ and $n_\mathrm{e}$ are the ion and electron number density, $\vector{j}_\mathrm{e}$ is the electron current density, $\vector{E}$ is the electric field, and $\vector{B}$ is the magnetic field \cite{Goebel2008_fundamentals}. Here, the Lorentz force exerted on ions can be neglected since the ion current density is small and the Lorentz force is sufficiently smaller than the electrostatic force. 

In the case of Hall thrusters, the electron current density $\vector{j}_\mathrm{e}$ that generates the Lorentz force is due to the $\vector{E} \times \vector{B}$ effect, i.e., the Hall current \cite{Goebel2008_fundamentals}. The electrostatic and Lorentz forces exerted on electrons are balanced, and the net force exerted on electrons $\vector{f}_\mathrm{e}$ is equal to zero theoretically. Therefore, electrons in Hall thrusters do not obtain the net momentum. Instead of electrons, only ions are accelerated by the electrostatic force and obtain the net axial momentum directed downstream. 

In the magnetic nozzle rf plasma thrusters, the Lorentz force is also exerted on electrons as well as Hall thrusters. However, the electron current density that produces the Lorentz force is due to the diamagnetic effect as reported in \cite{Takahashi2011_prl, Fruchtman2012_pop, Takahashi2013_prl, Takahashi2016_psst, Emoto2021_pop}. As a result, the force exerted on electrons in the magnetic nozzle is generally not zero because the electrostatic force $- e n_\mathrm{e} \vector{E}$ is not parallel to the Lorentz force $\vector{j}_\mathrm{e} \times \vector{B}$ and these forces are not canceled out. Therefore, electrons in the magnetic nozzle could obtain the net momentum by the electrostatic and Lorentz forces.

Previous studies showed that the Lorentz force in the magnetic nozzle increased with increasing the magnetic field strength \cite{Takahashi2016_psst, Emoto2021_pop}. When the magnetic field strength is sufficiently strong and the Lorentz force exceeds the electrostatic one ($|\vector{j}_\mathrm{e} \times \vector{B}| > |e n_\mathrm{e} \vector{E}|$), electrons in the magnetic nozzle would obtain the net momentum in the downstream direction by the Lorentz force while the electrostatic force decelerates electrons. In this situation, both ions and electrons could obtain the net momentum in the downstream direction. Previous thruster models have shown that the thrust corresponding to the total axial momentum of the ions and the electrons increases along the magnetic nozzle \cite{Fruchtman2006_prl, Takahashi2011_prl}. However, it has not been clearly revealed which gains the main axial momentum, ions and electrons.

Previous studies showed that a spontaneous electrostatic force did not give the net momentum to the plasma and just converted the electron pressure to the ion momentum \cite{Fruchtman2006_prl, Lafleur2010_pop}. However, it was also reported that the Lorentz force exerted on electrons in the magnetic nozzle imparted the net axial momentum to the plasma \cite{Takahashi2013_prl}. Although the detailed spatial measurement of the momentum change has been investigated in an experiment \cite{Takahashi2020_njp}, it has been still unclear how the momentums of ions and electrons are affected by the Lorentz force.

To investigate the ion and electron momentums in the magnetic nozzle, it is necessary to obtain the electrostatic and Lorentz forces independently and analyze momentum gains of ions and electrons given by these forces. In this study, we conduct particle-in-cell simulations of the magnetic nozzle acceleration with Monte Carlo collisions (PIC-MCCs), which treat the kinetics of both ions and electrons and an electrostatic field generated by these particles. Here, PIC-MCC simulations can investigate the momentum gains of ions and electrons independently while the previous model treated the electromagnetic thrust summed by ions and electrons \cite{Takahashi2011_prl}. Distributions of electrostatic and Lorentz forces and momentum gains per ion and electron are shown in section \ref{sec:results_and_discussion}. We clarify the dominant force in the magnetic nozzle and detect whether ions or electrons obtain the net axial momentum mainly. 

\section{Numerical model}
\label{sec:model}

We employ a two-dimensional and symmetric calculation model of a magnetic nozzle rf plasma thruster to avoid the central anomaly and reduce the calculation cost. This symmetric model simulates the bi-directional thruster, which is expected to be used for space debris removal \cite{Takahashi2018_sr}. Even in the bi-directional configuration, the essential plasma dynamics in the magnetic nozzle is not affected by such a symmetric configuration.

\begin{figure}
    \centering
    \includegraphics{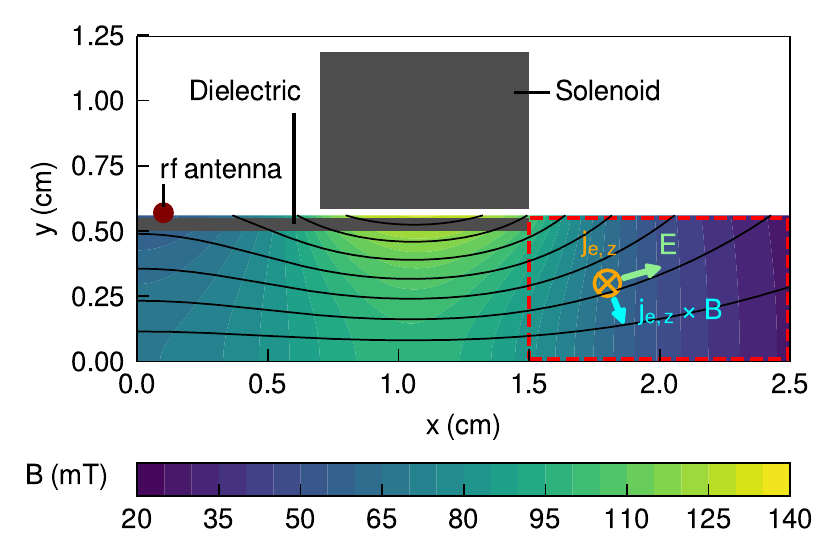}
    \caption{A schematic of the calculation area and a magnetic field strength produced by the solenoid at $I_B$ = 2.0 kA turn. Solid black lines show the magnetic field lines. Subsequent results are shown within a red-dotted rectangle area although the entire calculation area is 2.5 cm $\times$ 0.56 cm.}
    \label{fig:solenoid_magnetic_field}
\end{figure}

Figure \ref{fig:solenoid_magnetic_field} shows a schematic of the calculation area employed in the PIC-MCC simulation. The calculation area is 2.5 cm $\times$ 0.56 cm including the dielectric. The calculation model is roughly one-sixth size of the experiment and consists of a radiofrequency (rf) antenna, a dielectric, and a solenoid \cite{Takahashi2018_sr}. The solenoid produces the magnetic nozzle by the magnetostatic field, which accelerates the plasma generated by the inductively coupled mode. The electrostatic field $\vector{E}_{es}$ and the Lorentz force $\vector{j}_{\mathrm{e},z} \times \vector{B}$ are exerted on the plasma in the magnetic nozzle. Subsequent results in section \ref{sec:results_and_discussion} are shown within a red-dotted rectangle area in figure \ref{fig:solenoid_magnetic_field} to investigate the momentum gain processes in the magnetic nozzle. 

The PIC-MCC simulation consists of the kinetics of charged particles, an electrostatic field $\vector{E}_\mathrm{es}$, the electromagnetic field induced by the rf antenna $\vector{E}_\mathrm{em}$, and the magnetostatic field produced by the solenoid $\vector{B}$. The detail of the PIC-MCC simulation has been described in our previous papers so that we briefly describe the numerical model in this paper \cite{Takao2015_pop, Takase2018_pop,Emoto2021_pop}. Here, our previous study showed that the PIC-MCC model well reproduced the plasma profiles and the diamagnetic effect measured in previous experiments \cite{Emoto2021_pop}. The equations of motions of charged particles are solved by using the Boris method \cite{Birdsall2004_crc}. In this PIC-MCC simulation, we calculate elastic, excitation, and ionization collisions using null-collision method \cite{Vahedi1995_cpc}. The electrostatic field $\vector{E}_\mathrm{es}$ is obtained from the Poisson's equation by using the fast Fourier transformation with the Dirichlet boundary condition of $\phi$ = 0 at $x$ = 2.5 cm and $y$ = 0.56 cm, where $\phi$ is the potential. The electric field induced by the rf antenna $\vector{E}_\mathrm{em}$ is obtained from Maxwell's equations by using the fast Fourier transformation with the Dirichlet boundary condition of $\vector{E}_\mathrm{em}$ = 0 at $x$ = 2.5 cm and $y$ = 0.56 cm. The solenoid magnetic field $\vector{B}$ is obtained from Maxwell's equations by using the fast Fourier transformation in ten times larger calculation area. Here, the solenoid current is set to 0, 0.4, and 2.0 kA turn to investigate the dependence of the magnetic field strength. Note that the solenoid current of 0 kA turn means no magnetic field in comparison with the magnetic nozzle acceleration. Figure \ref{fig:solenoid_magnetic_field} also shows the solenoid magnetic field strength for the solenoid current of 2.0 kA turn as a colormap and the magnetic field lines as solid black lines.

Table \ref{tab:calculation_condition} shows the summary of calculation conditions employed in our PIC-MCC simulation. The calculation area is divided into 50 $\mu$m $\times$ 50 $\mu$m cells. We put a hundred ions and electrons per cell as initial particles, respectively. The rf frequency is set to 80 MHz, and the rf current is controlled to satisfy that the power absorption by the charged particle is 3.5 W/m. We treat singly charged xenon ions Xe$^+$ and electrons e$^-$ as charged particles. The time step of ions is set to  0.125 ns as 1/100 rf period, and that of electrons is set to 3.57 ps as 1/35 time step for ions. Neutrals are set to constant spatiotemporally as the neutral density of 2.0 $\times$ 10$^{19}$ m$^{-3}$ and the neutral temperature of 300 K. For the above-mentioned numerical configurations, the radial ($y$) profile of the plasma density is bi-modal for the strong magnetic field case as observed in an earlier simulation \cite{Emoto2021_pop} and in an experiment \cite{Takahashi2017_pop}, while it has a central peak for a weak magnetic field.

\begin{table}
    \centering
    \caption{Calculation conditions.}
    \label{tab:calculation_condition}
    \begin{tabular}{ll}
        \hline
        Cell size & 50 $\mu$m \\
        rf frequency & 80 MHz \\
        Power absorption & 3.5 W/m \\
        Particle & Xe$^+$ and e$^-$ \\
        Time step for ions & 0.125 ns (1/100 rf period) \\
        Time step for electrons & 3.57 ps (1/35 time step for ions) \\
        Neutral density & 2.0 $\times$ 10$^{19}$ m$^{-3}$ \\
        Neutral temperature & 300 K \\
        Electron-neutral collisions & Elastic, excitation, and ionization \\
        \hline
    \end{tabular}
\end{table}

The electrostatic force $f_{\mathrm{E},x}$ and the Lorentz force $f_{\mathrm{L},x}$ exerted on an electron in the $x$-direction are written as

\begin{equation}
    \label{eq:electrostatic_force}
    f_{\mathrm{E},x} = - e E_{\mathrm{es},x},
\end{equation}

\begin{equation}
    \label{eq:lorentz_force}
    f_{\mathrm{L},x} = e B_y \frac{\sum_k v_{\mathrm{e},k,z}}{N},
\end{equation}

\noindent
where $k$ is the index of particles, $N$ is the number of particles in a cell, and $v_{\mathrm{e},k,z}$ is the velocity of electron $k$ in the $z$-direction, being equivalent to the azimuthal direction in a cylindrical coordinate. Here, the electrostatic force exerted on a singly charged ion is written as $- f_{\mathrm{E},x} = e E_{\mathrm{es},x}$ because the Lorentz force exerted on ions can be negligible as mentioned in section \ref{sec:introduction}.

The momentums of the charged particles in the magnetic nozzle are changed by the electrostatic and Lorentz forces in equations \eqref{eq:electrostatic_force} and \eqref{eq:lorentz_force}. Here, we define momentum gains per particle in the $x$-direction $\Delta \dot{M}_x$ and $y$-direction $\Delta \dot{M}_y$ as

\begin{equation}
    \label{eq:momentum_gain_per_particle_1}
    \Delta \dot{M}_x = \frac{\sum_k \Delta (m_k v_{k,x})}{N \Delta t},
\end{equation}

\begin{equation}
    \label{eq:momentum_gain_per_particle_2}
    \Delta \dot{M}_y = \frac{\sum_k \Delta (m_k v_{k,y})}{N \Delta t},
\end{equation}

\noindent
respectively, where $\Delta t$ is the time step and $\Delta (m_k v_{k,x})$ is the momentum gain of the particle $k$ in the $x$-direction for time step $\Delta t$. The momentum gain per particle is averaged over 30 $\mu$s. In addition, we define the net momentum gains per particle in the $x$-direction $\Delta \dot{M}_{\mathrm{net},x}$ and the $y$-direction $\Delta \dot{M}_{\mathrm{net},y}$ as

\begin{equation}
    \label{eq:net_momentum_1}
    \Delta \dot{M}_{\mathrm{net},x} = \Delta \dot{M}_{\mathrm{i},x} + \Delta \dot{M}_{\mathrm{e},x},
\end{equation}

\begin{equation}
    \label{eq:net_momentum_2}
    \Delta \dot{M}_{\mathrm{net},y} = \Delta \dot{M}_{\mathrm{i},y} + \Delta \dot{M}_{\mathrm{e},y},
\end{equation}

\noindent
respectively, where $\Delta \dot{M}_{\mathrm{i},x}$ and $\Delta \dot{M}_{\mathrm{e},x}$ are the momentum gains per ion and electron in the $x$-direction and $\Delta \dot{M}_{\mathrm{i},y}$ and $\Delta \dot{M}_{\mathrm{e},y}$ are those in the $y$-direction, calculated from equations \eqref{eq:momentum_gain_per_particle_1} and \eqref{eq:momentum_gain_per_particle_2}.

\section{Results and discussion}
\label{sec:results_and_discussion}

\begin{figure}
    \centering
    \includegraphics[width=.95\linewidth]{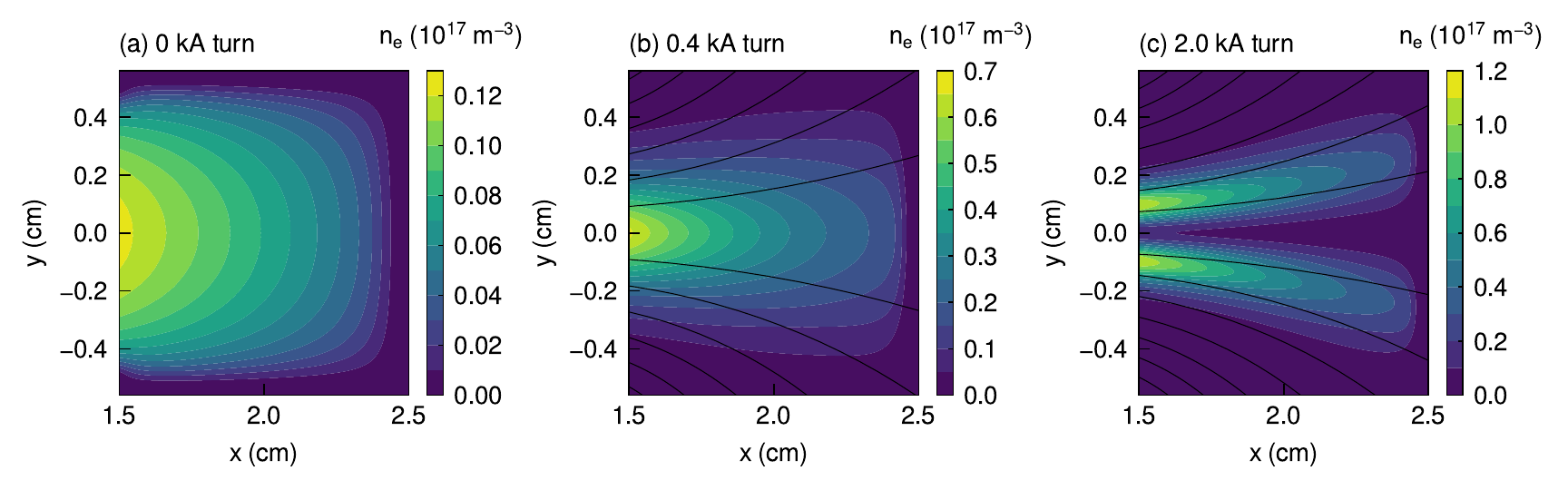}
    \caption{$x$-$y$ profiles of the electron number density for the three solenoid currents of (a) 0, (b) 0.4, and (c) 2.0 kA turn. Solid black lines show the magnetic field lines produced by the solenoid.}
    \label{fig:electron_number_density}
\end{figure}

Figure \ref{fig:electron_number_density} shows $x$-$y$ profiles of the electron number density in the downstream region of the source for the three solenoid currents of 0, 0.4, and 2.0 kA turn. The electron number density increases with increasing the solenoid current since the electrons are well confined by the magnetic field and less likely to be lost to the dielectric wall and calculation boundaries. In addition, the $y$-profile of the electron number density for the solenoid current of 2.0 kA turn shows the bi-modal shape, which was measured in previous experiments \cite{Takahashi2009_apl, Charles2010_apl, Ghosh2017_pop, Gulbrandsen2017_fr, Takahashi2017_pop}.

\begin{figure}
    \centering
    \includegraphics{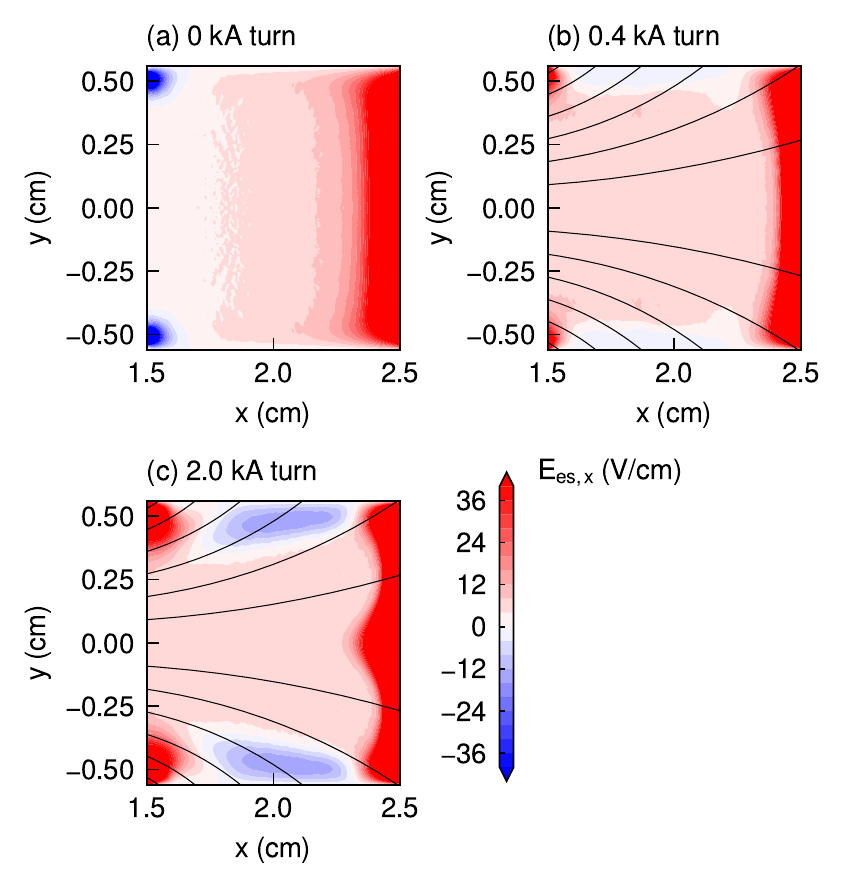}
    \caption{$x$-$y$ profiles of the electrostatic field in the $x$-direction $E_{\mathrm{es},x}$ for the three solenoid currents of (a) 0, (b) 0.4, and (c) 2.0 kA turn. Solid black lines show the magnetic field lines produced by the solenoid.}
    \label{fig:electrostatic_field_1}
\end{figure}

\begin{figure}
    \centering
    \includegraphics{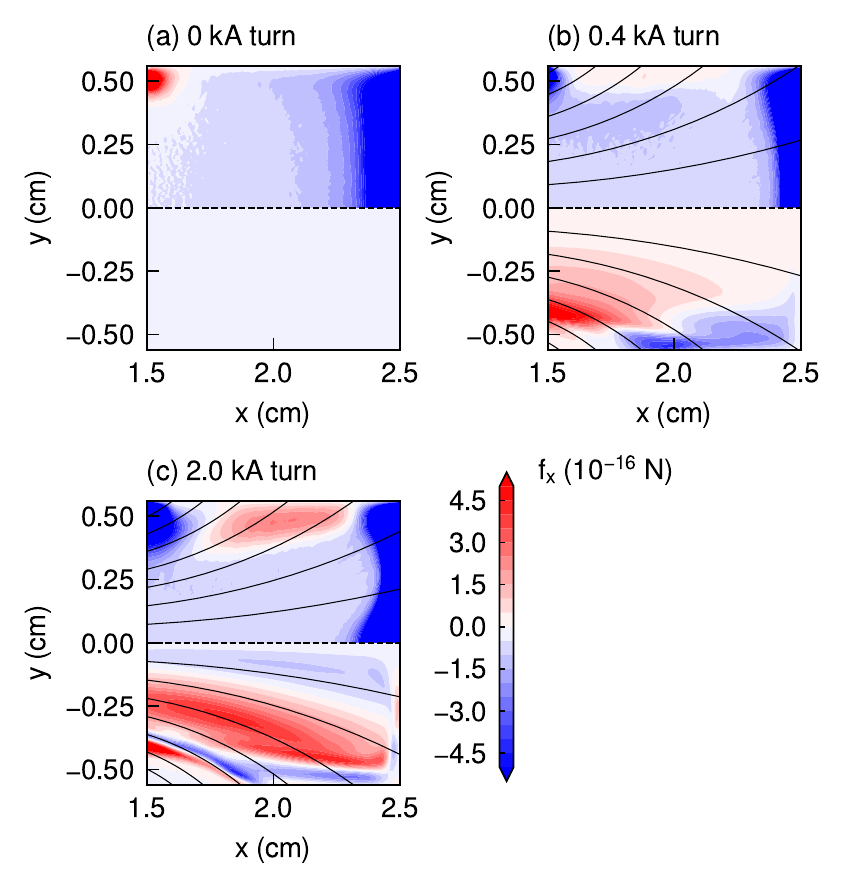}
    \caption{$x$-$y$ profiles of the electrostatic force $f_{\mathrm{E},x}$ (upper half) and the Lorentz force $f_{\mathrm{L},x}$ (lower half) exerted on an electron in the $x$-direction for the three solenoid currents of (a) 0, (b) 0.4, and (c) 2.0 kA turn. Solid black lines show the magnetic field lines produced by the solenoid.}
    \label{fig:electrostatic_and_lorentz_force_1}
\end{figure}

Figure \ref{fig:electrostatic_field_1} shows $x$-$y$ profiles of the electrostatic field in the $x$-direction $E_{\mathrm{es},x}$ for the three solenoid currents of 0, 0.4, and 2.0 kA turn. The electrostatic field in the $x$-direction $E_{\mathrm{es},x}$ is positive in the almost entire region since the plasma expands through the magnetic nozzle and the potential decreases in the downstream direction. It should be noted that the negative electrostatic field is observed at the peripheral region in $x$ = 1.7--2.3 cm and $y$ = $\pm$(0.3--0.56) cm and decreases with increasing the solenoid current, especially for the 2.0 kA turn solenoid current case. This negative electrostatic field implies that the high potential is formed in the plasma core flowing along the divergent magnetic field lines. The electron number density in approximately $|y|$ > 0.3 cm decreases directed to the negative $x$-direction as shown in figures \ref{fig:electron_number_density}(b) and \ref{fig:electron_number_density}(c), resulting in the potential gradient directed in the positive $x$-direction and the negative electrostatic field.

Figure \ref{fig:electrostatic_and_lorentz_force_1} shows $x$-$y$ profiles of the electrostatic force $f_{\mathrm{E},x}$ (upper half) and the Lorentz force $f_{\mathrm{L},x}$ (lower half) exerted on an electron in the $x$-direction for the three solenoid currents of 0, 0.4, and 2.0 kA turn. It should be noted that the sheath is generated at $x$ = 2.3--2.5 cm and $y$ = $\pm$(0.4--0.56) cm since the electrostatic field $\vector{E}_\mathrm{es}$ is solved with the Dirichlet boundary condition as $\phi$ = 0. Here, the Lorentz force for the solenoid current of 0 kA turn is zero as seen in figure \ref{fig:electrostatic_and_lorentz_force_1}(a) because of the absence of the magnetic field. 

The electrostatic force exerted on an electron $f_{\mathrm{E},x}$ is negative in the almost entire region, and its magnitude is not significantly changed by the solenoid current. As a result, the electrostatic force $f_{\mathrm{E},x}$ decelerates electrons expanding through the magnetic nozzle and accelerates ions in the downstream direction instead. The Lorentz force exerted on an electron $f_{\mathrm{L},x}$ increases dramatically with increasing the solenoid current, exceeding the electrostatic force $f_{\mathrm{E},x}$ clearly for the solenoid current of 2.0 kA turn. The negative Lorentz force also exists at the center of the magnetic nozzle in figure \ref{fig:electrostatic_and_lorentz_force_1}(c), which is due to the bi-modal plasma shape as shown in figure \ref{fig:electron_number_density}(c). The positive Lorentz force would accelerate electrons in the downstream direction while the electrostatic force decelerates them.

\begin{figure}
    \centering
    \includegraphics{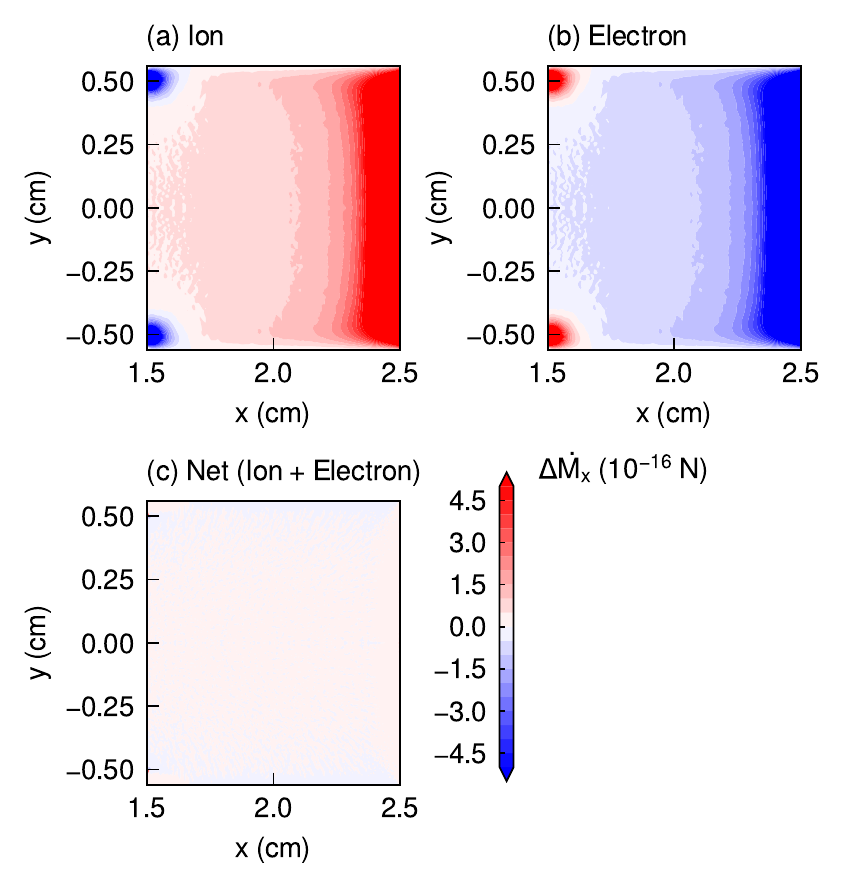}
    \caption{$x$-$y$ profiles of the axial momentum gains per (a) ion $\Delta \dot{M}_{\mathrm{i},x}$ and (b) electron $\Delta \dot{M}_{\mathrm{e},x}$ for the solenoid current of 0 kA turn. The net momentum gain per particle $\Delta \dot{M}_{\mathrm{net},x}$ is also shown in (c), which is the sum of $\Delta \dot{M}_{\mathrm{i},x}$ and $\Delta \dot{M}_{\mathrm{e},x}$.}
    \label{fig:momentum_gain_per_particle_1_00}
\end{figure}

\begin{figure}
    \centering
    \includegraphics{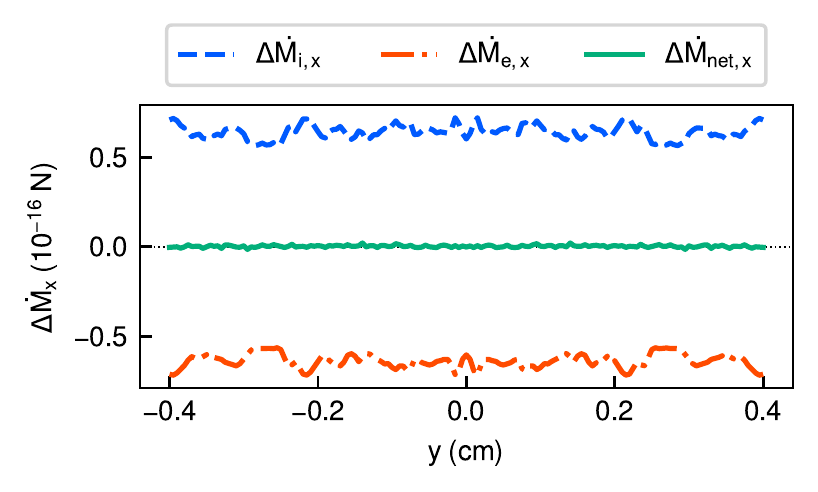}
    \caption{$y$ profile of the axial momentum gain per ion $\Delta \dot{M}_{\mathrm{i},x}$ (a dashed blue line) and electron $\Delta \dot{M}_{\mathrm{e},x}$ (a dotted-dashed orange line) for the solenoid current of 0 kA turn at $x$ = 1.8 cm. The net momentum gain per particle $\Delta \dot{M}_{\mathrm{net},x}$ (a solid green line) is also plotted, which is the sum of $\Delta \dot{M}_{\mathrm{i},x}$ and $\Delta \dot{M}_{\mathrm{e},x}$. The data for $|y|$ > 0.4 cm is eliminated because it is affected by the sheath due to the finite calculation area.}
    \label{fig:section_momentum_gain_per_particle_1_00}
\end{figure}

Figure \ref{fig:momentum_gain_per_particle_1_00} shows $x$-$y$ profiles of the axial momentum gains per ion $\Delta \dot{M}_{\mathrm{i},x}$ and electron $\Delta \dot{M}_{\mathrm{e},x}$ for the solenoid current of 0 kA turn, which are directly calculated from the velocities of the ions and the electrons as described in equation \eqref{eq:momentum_gain_per_particle_1}. The net axial momentum gain per particle $\Delta \dot{M}_{\mathrm{net},x}$ is also calculated, which is the sum of $\Delta \dot{M}_{\mathrm{i},x}$ and $\Delta \dot{M}_{\mathrm{e},x}$ as shown in equation \eqref{eq:net_momentum_1}. It should be noted that the magnitudes of the axial momentum gains per ion and electron are large in $x$ = 2.3--2.5 cm because of the sheath. The axial momentum gain per ion $\Delta \dot{M}_{\mathrm{i},x}$ is positive in the almost entire region, indicating that ions obtain the momentum in the $x$-direction and are accelerated in the downstream direction by the electrostatic field. However, the axial momentum gain per electron $\Delta \dot{M}_{\mathrm{e},x}$ is negative in almost all regions. Whereas low-energy electrons are reflected by the sheath, energetic electrons are decelerated by the electrostatic field and lose the momentum at the boundary. The net axial momentum gain per particle $\Delta \dot{M}_{\mathrm{net},x}$ is almost zero as shown in figure \ref{fig:momentum_gain_per_particle_1_00}(c) so that the axial momentum gains per ion and electron are canceled out each other. This result is consistent with the analytical prediction \cite{Fruchtman2006_prl} and the experiments \cite{Takahashi2011_apl, Lafleur2011_pop}. Figure \ref{fig:section_momentum_gain_per_particle_1_00} shows $y$ profile of the axial momentum gains per particle for the solenoid current of 0 kA turn at $x$ = 1.8 cm. The axial momentum gains per ion and electron are symmetric, and the net axial momentum gain per particle $\Delta \dot{M}_{\mathrm{net},x}$ shows completely zero. Therefore, our PIC-MCC simulation demonstrates that the plasma without the magnetic field does not obtain the net momentum as reported in \cite{Fruchtman2006_prl}. 

\begin{figure}
    \centering
    \includegraphics{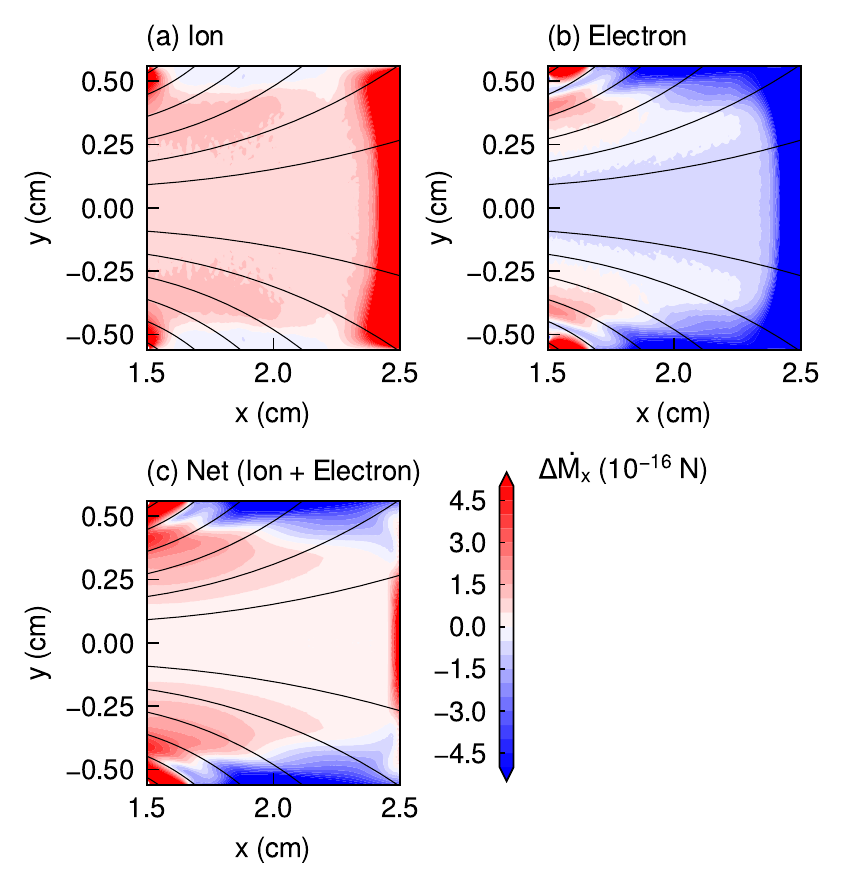}
    \caption{$x$-$y$ profiles of the axial momentum gains per (a) ion $\Delta \dot{M}_{\mathrm{i},x}$ and (b) electron $\Delta \dot{M}_{\mathrm{e},x}$ for the solenoid current of 0.4 kA turn. The net axial momentum gain per particle $\Delta \dot{M}_{\mathrm{net},x}$ is also shown in (c), which is the sum of $\Delta \dot{M}_{\mathrm{i},x}$ and $\Delta \dot{M}_{\mathrm{e},x}$. Solid black lines show the magnetic field lines produced by the solenoid.}
    \label{fig:momentum_gain_per_particle_1_04}
\end{figure}

\begin{figure}
    \centering
    \includegraphics{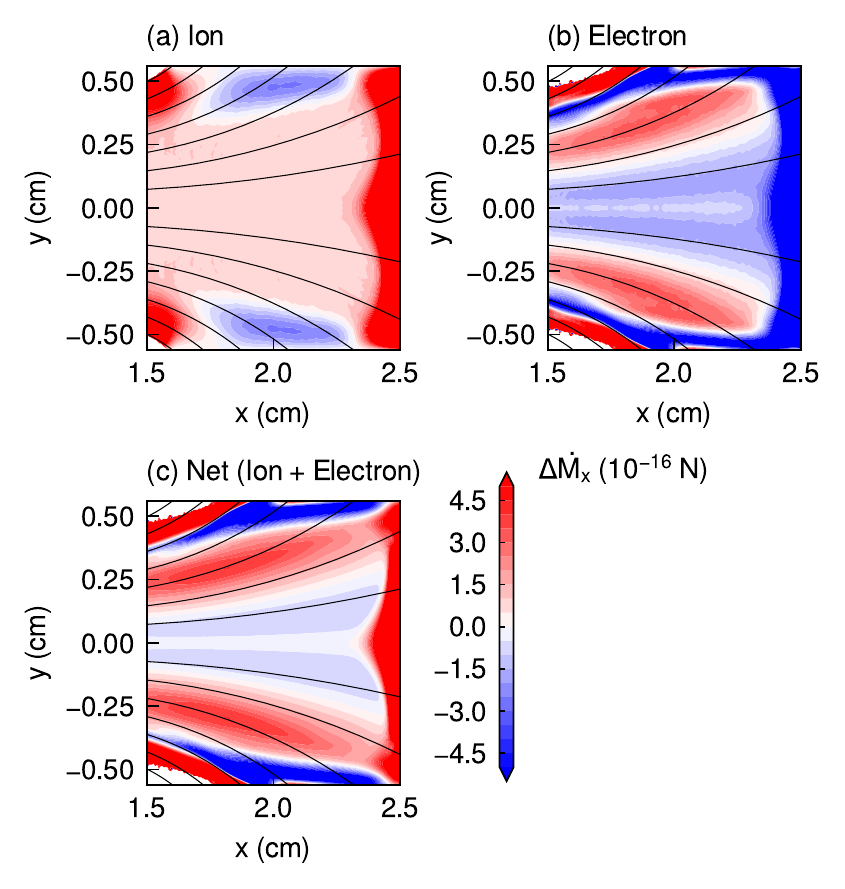}
    \caption{$x$-$y$ profiles of the axial momentum gains per (a) ion $\Delta \dot{M}_{\mathrm{i},x}$ and (b) electron $\Delta \dot{M}_{\mathrm{e},x}$ for the solenoid current of 2.0 kA turn. The net axial momentum gain per particle $\Delta \dot{M}_{\mathrm{net},x}$ is also shown in (c), which is the sum of $\Delta \dot{M}_{\mathrm{i},x}$ and $\Delta \dot{M}_{\mathrm{e},x}$. Solid black lines show the magnetic field lines produced by the solenoid.}
    \label{fig:momentum_gain_per_particle_1_20}
\end{figure}

Figures \ref{fig:momentum_gain_per_particle_1_04} and \ref{fig:momentum_gain_per_particle_1_20} show $x$-$y$ profiles of the axial momentum gains per ion $\Delta \dot{M}_{\mathrm{i},x}$ and electron $\Delta \dot{M}_{\mathrm{e},x}$ for the two solenoid currents of 0.4 and 2.0 kA turn, respectively. The net axial momentum gain per particle $\Delta \dot{M}_{\mathrm{net},x}$ is also calculated, which is the sum of $\Delta \dot{M}_{\mathrm{i},x}$ and $\Delta \dot{M}_{\mathrm{e},x}$ as shown in equation \eqref{eq:net_momentum_1}. It should be noted that the region in $x$ = 2.4--2.5 cm and $y$ = $\pm$(0.4--0.56) cm contains the effect of the sheath near the boundaries. The axial momentum gain per ion $\Delta \dot{M}_{\mathrm{i},x}$ is positive in the almost entire region regardless for both of the solenoid currents as shown in figures \ref{fig:momentum_gain_per_particle_1_04}(a) and \ref{fig:momentum_gain_per_particle_1_20}(a) so that ions obtain the axial momentum and are accelerated by the electrostatic force. The axial momentum gain per electron $\Delta \dot{M}_{\mathrm{e},x}$ is almost negative for the solenoid current of 0.4 kA turn, while the positive axial momentum gain per electron also exists on the magnetic field line passing through $x$ = 1.5 cm and $y$ = $\pm$0.35 cm, indicating that the electron momentum increases by the Lorentz force and exceeds the electrostatic force around there. For the solenoid current of 2.0 kA turn, the positive axial momentum gain per electron clearly exists on the magnetic field line passing through $x$ = 1.5 cm and $y$ = $\pm$0.25 cm whereas the negative axial momentum gain per electron also exists at the center of the magnetic nozzle. Regions where the positive axial momentum gain per electron exists are consistent with the locations where the Lorentz force in the $x$-direction is exerted on electrons in figure \ref{fig:electrostatic_and_lorentz_force_1}(c). In this region, the Lorentz force exceeds the electrostatic one and accelerates electrons in the downstream direction even though the electrostatic force decelerates electrons. At the center of the magnetic nozzle, the negative axial momentum gain per electron exists for the solenoid current of 2.0 kA turn since both the electrostatic and Lorentz forces exerted on electrons are negative as shown in figure \ref{fig:electrostatic_and_lorentz_force_1}(c). In this region, electrons are decelerated by both the electrostatic and Lorentz forces.

As shown in figures \ref{fig:momentum_gain_per_particle_1_04}(c) and \ref{fig:momentum_gain_per_particle_1_20}(c), the net axial momentum gain per particle $\Delta \dot{M}_{\mathrm{net},x}$ is not zero unlike that without the magnetic field in figure \ref{fig:momentum_gain_per_particle_1_00}(c). It is consistent with the previous study as reported in \cite{Takahashi2013_prl} that the magnetic nozzle imparts the net momentum to the plasma. In this situation, the plasma momentum is not only converted by the spontaneous electric field but also the Lorentz force with the magnetic field. For the solenoid current of 0.4 kA turn, the positive net axial momentum gain per particle is because the axial momentum gain per electron increases by the Lorentz force and that per ion becomes dominant. For the solenoid current of 2.0 kA turn, however, the axial momentum gain per electron increases significantly by the Lorentz force and becomes dominant instead of ions. These results indicate that the electron momentum imparted by the strong magnetic field exceeds the ion momentum imparted by the electrostatic field. Here, it is still unclear that the net axial momentum gain per particle near the right boundary at $x$ = 2.5 cm is large, where the sheath is generated because of the boundary condition. One of the possible reasons would be the generation of the $\vector{E} \times \vector{B}$ drift current due to the presence of the strong electric field in the sheath and further verification will be required to understand it fully. Since the present paper focuses on the momentum gain in the core region (not in the sheath), this is out of scope in the present paper. 

\begin{figure}
    \centering
    \includegraphics{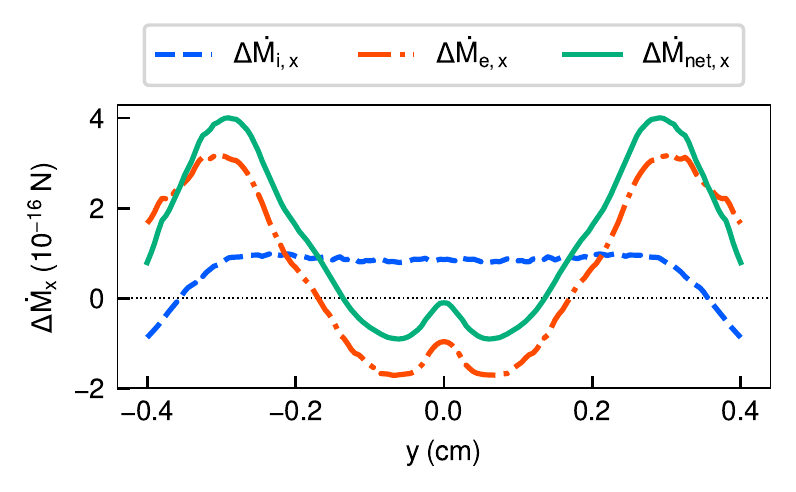}
    \caption{$y$ profile of the axial momentum gain per ion $\Delta \dot{M}_{\mathrm{i},x}$ (a dashed blue line) and electron $\Delta \dot{M}_{\mathrm{e},x}$ (a dotted-dashed orange line) for the solenoid current of 2.0 kA turn at $x$ = 1.8 cm. The net axial momentum gain per particle $\Delta \dot{M}_{\mathrm{net},x}$ (a solid green line) is also plotted, which is the sum of $\Delta \dot{M}_{\mathrm{i},x}$ and $\Delta \dot{M}_{\mathrm{e},x}$. The data for $|y|$ > 0.4 cm is eliminated because it is affected by the sheath due to the finite calculation area.}
    \label{fig:section_momentum_gain_per_particle_1_20}
\end{figure}

Figure \ref{fig:section_momentum_gain_per_particle_1_20} shows $y$ profile of the axial momentum gains per particle for the solenoid current of 2.0 kA turn at $x$ = 1.8 cm. Note that the axial momentum gains per particle in $|y|$ > 0.4 cm are eliminated because they are affected by the sheath. The axial momentum gain per ion is positive and uniform at approximately 1.0 $\times$ 10$^{-16}$ N between $-$0.3 cm < $y$ < 0.3 cm. The axial momentum gain per electron $\Delta \dot{M}_{\mathrm{e},x}$ is larger than that per ion in outer regions of $|y|$ > 0.2 cm, where the electrons are accelerated by the Lorentz force in the downstream direction instead of ions. However, the electrostatic force exceeds the Lorentz force in the inner region of $|y|$ < 0.2 cm and accelerates ions in the downstream direction instead of electrons. Here, the net axial momentum gain per particle is dominated by the axial momentum gain per electron in outer regions of $|y|$ > 0.2 cm at the strong magnetic field. 

The positive axial momentum gain per electron in figure \ref{fig:momentum_gain_per_particle_1_20}(b) implies that the electron energy in the $x$-direction increases by the magnetic nozzle. However, the magnetostatic field produced by the solenoid does not give the energy to the electrons. Therefore, the increase of the electron energy in the $x$-direction must be from the internal energy of the plasma. 

The spontaneous electrostatic field converts the electron pressure in the $x$-directions to the ion momentum in the $x$-direction as reported in \cite{Fruchtman2006_prl}. However, the spontaneous electrostatic field does not convert the electron pressure in the $y$- and $z$-directions to the ion momentum in the $x$-direction. To increase the axial electron momentum without the external work, the energy corresponding to the electron pressure in the $y$- and $z$-directions should be converted to the $x$-direction by the Lorentz force.

\begin{figure}
    \centering
    \includegraphics{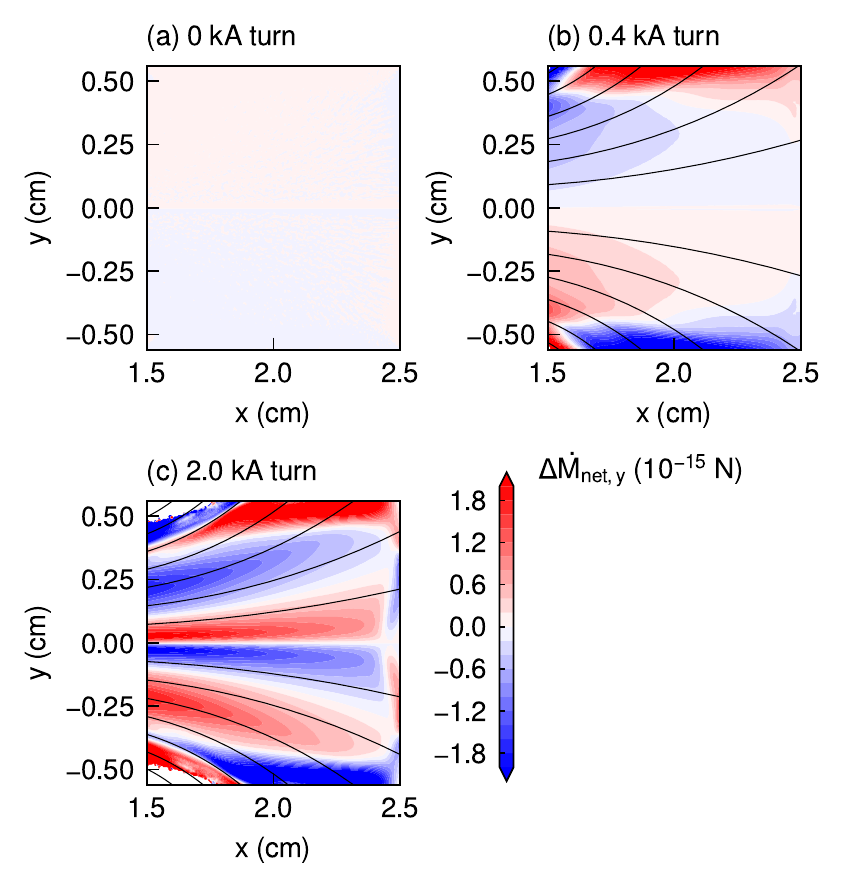}
    \caption{$x$-$y$ profiles of the net momentum gain per particle in the $y$-direction $\Delta \dot{M}_{\mathrm{net},y}$ for the solenoid currents of (a) 0, (b) 0.4, and (c) 2.0 kA turn. Solid black lines show the magnetic field lines produced by the solenoid.}
    \label{fig:momentum_gain_per_particle_2}
\end{figure}

Figure \ref{fig:momentum_gain_per_particle_2} shows $x$-$y$ profiles of the net momentum gain per particle in the $y$-direction $\Delta \dot{M}_{\mathrm{net},y}$ for the solenoid currents of 0, 0.4, and 2.0 kA turn. Comparing figure \ref{fig:momentum_gain_per_particle_2} with figures \ref{fig:momentum_gain_per_particle_1_00}(c), \ref{fig:momentum_gain_per_particle_1_04}(c), and \ref{fig:momentum_gain_per_particle_1_20}(c), we can see that the regions where the net momentum gain per particle in the $y$-direction increases or decreases correspond to those where the net momentum gain per particle in the $x$-direction increases. Here, the net momentum gain is calculated as the sum of the ion and electron momentums, and the effect of the electrostatic field is canceled out, indicating the momentum conversion by the Lorentz force. Thus, the role of the magnetic nozzle is converting the electron momentum in the $y$-direction into that in the $x$-direction to utilize the electron energy efficiently. Note that the net momentum gain per particle in the $z$-direction is confirmed to be neglected because it is calculated to be approximately two orders of magnitude smaller than that in the $x$- and $y$-directions.

\begin{figure}
    \centering
    \includegraphics{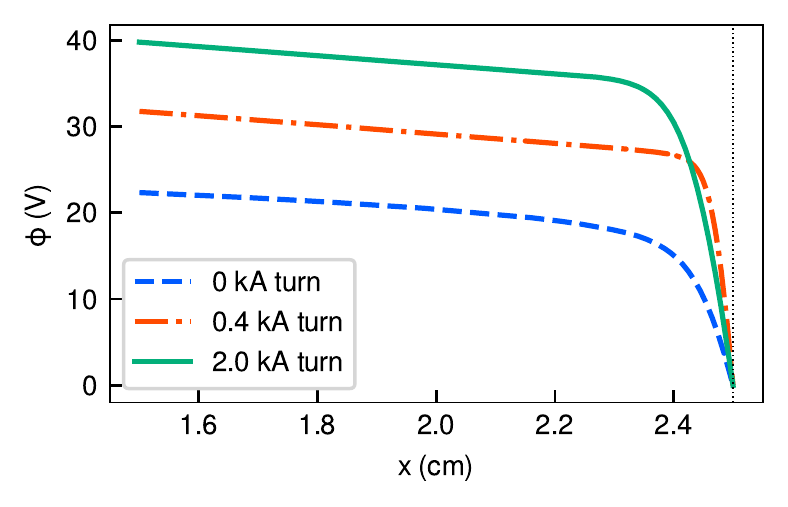}
    \caption{$x$ profile of the potential $\phi$ for the three solenoid currents of 0 (a dashed blue line), 0.4 (a dotted-dashed orange line), and 2.0 kA turn (a solid green line) at $y$ = 0 cm. A vertical dotted black line at $x$ = 2.5 cm shows the right boundary of the calculation area, where the Dirichlet boundary condition of $\phi$ = 0 is assumed.}
    \label{fig:section_scalar_potential}
\end{figure}

The positive axial momentum gain per electron given by the Lorentz force means that electrons are accelerated to the downstream direction and are more likely to be lost to the downstream boundary. In this situation, the plasma potential would increase to prevent the loss of electrons. Figure \ref{fig:section_scalar_potential} shows $x$ profile of the potential for the three solenoid currents of 0, 0.4, and 2.0 kA turn at $y$ = 0 cm. The plasma potential on the left side in figure \ref{fig:section_scalar_potential} increases with increasing the solenoid current, indicating that the plasma prevents the loss of electrons spontaneously. In addition, the increase of the plasma potential further accelerates ions by the electrostatic field, resulting in the increase of the exhaust velocity.

To investigate the net thrust obtained by the magnetic nozzle, we calculate the total axial momentum gain by integrating the ion and electron momentum gains in the $x$-direction in 1.5 cm < $x$ < 2.3 cm and $-$0.35 cm < $y$ < 0.35 cm eliminating the sheath effect. Table \ref{tab:total_momentum_gain} shows total axial momentum gain in the magnetic nozzle. For the solenoid current of 0 kA turn, the magnitude of the axial momentum gains of ions and electrons are almost the same, indicating that the net axial momentum gain becomes small. For the solenoid current of 0.4 kA turn, the axial momentum gains of ions and electrons increase, and the net axial momentum gain also increases. For the solenoid current of 2.0 kA turn, the axial momentum gain of electrons increases dramatically while the axial momentum gain of ions remains almost unchanged. As a result, the axial momentum gain of electrons becomes positive and exceeds that of ions. The net axial momentum gain also increases significantly though the negative net axial momentum gain exists in the center of the magnetic nozzle as shown in figure \ref{fig:momentum_gain_per_particle_1_20}(c). Here, the electron momentum is finally converted to the ion momentum in the sheath so that the strong magnetic field also increases the ion exhaust velocity.

\begin{table}
    \centering
    \caption{Total axial momentum gain in the magnetic nozzle.}
    \label{tab:total_momentum_gain}
    \begin{tabular}{llll}
        \hline
         Solenoid current (kA turn) &  \multicolumn{3}{l}{Axial Momentum gain ($\mu$N/m)} \\
         & Ion & Electron & Net (Ion + Electron) \\
         \hline
         0   & 16.8 & $-$17.9 & $-$1.13 \\
         0.4 & 37.1 & $-$9.64 &    27.5 \\
         2.0 & 34.6 &    47.1 &    81.7 \\
         \hline
    \end{tabular}
\end{table}

These results are different from the phenomena in Hall thrusters, where the axial momentum gain per electron is zero theoretically. In the magnetic nozzle, the axial momentum gain per electron is not zero because of the Lorentz force due to the diamagnetic effect, and electrons obtain the net axial momentum contributing to the increase in the thrust and the exhaust velocity. The magnetic nozzle has the mechanism to obtain the thrust by accelerating electrons, where the electron momentum in the $y$-direction is converted into that in the $x$-direction.

\section{Conclusion}

We have conducted the particle-in-cell simulations of the bi-directional magnetic nozzle rf plasma thruster with Monte Carlo collisions to investigate the axial momentum gains of ions and electrons in the magnetic nozzle. The axial momentum gains per ion and electron are calculated directly from particle velocities, and the results are discussed with the calculated electrostatic and Lorentz forces, which are exerted on ions and electrons and impart the momentum. The Lorentz force in the $x$-direction increases with increasing the solenoid current and exceeds the electrostatic force in the $x$-direction at the strong magnetic field strength. The axial momentum gain per electron is also increased dramatically by the Lorentz force and becomes dominant in the magnetic nozzle instead of ions. It is clearly shown that the increase of the electron momentum in the $x$-direction is due to the momentum conversion of electrons from the $y$- to $x$-direction by the Lorentz force. The plasma potential also increases because of the loss of electrons, resulting in the increase in the exhaust velocity of ions. Therefore, the magnetic nozzle obtains the thrust by mainly imparting the net momentum in the $x$-direction to electrons.

\section*{Acknowledgement}

This work was partly supported by JSPS KAKENHI Grant Number JP19H00663. The computer simulation was performed on the A-KDK computer system at Research Institute for Sustainable Humanosphere, Kyoto University. One of the authors (K.E.) received a scholarship from The Futaba Foundation.

\section*{Data Availability}

The data that support the findings of this study are available upon reasonable request from the authors.

\bibliographystyle{iopart-num}
\bibliography{main}

\end{document}